\documentclass[prb,aps,showpacs,twocolumn,unsortedaddress,superscriptaddress]{revtex4}
\usepackage{graphics,bm}
\usepackage{epsfig,epsf}
\usepackage{graphicx,amsfonts,amsbsy,amssymb}

\begin{document}

\title{Inelastic scattering in ferromagnetic and antiferromagnetic metal spintronics}

\author{R.A. Duine}
\email{duine@phys.uu.nl} \homepage{http://www.phys.uu.nl/~duine}

\affiliation{Institute for Theoretical Physics, Utrecht
University, Leuvenlaan 4, 3584 CE Utrecht, The Netherlands}
\date{\today}

\affiliation{The University of Texas at Austin, Department of
Physics, 1 University Station C1600, Austin, TX 78712-0264}

\author{P.M. Haney}
\email{haney411@physics.utexas.edu}
\homepage{http://www.ph.utexas.edu/~haney411/paulh.html}

\affiliation{The University of Texas at Austin, Department of
Physics, 1 University Station C1600, Austin, TX 78712-0264}

\author{A. S. N\'u\~nez}
\email{alvaro.nunez@ucv.cl}
\homepage{http://www.ph.utexas.edu/~alnunez}

\affiliation{ Instituto de F\'isica, PUCV Av. Brasil 2950,
Valpara\'iso, Chile}

\affiliation{The University of Texas at Austin, Department of
Physics, 1 University Station C1600, Austin, TX 78712-0264}

\author{A.H. MacDonald}
\email{macd@physics.utexas.edu}
\homepage{http://www.ph.utexas.edu/~macdgrp}

\affiliation{The University of Texas at Austin, Department of
Physics, 1 University Station C1600, Austin, TX 78712-0264}

\date{\today}

\begin{abstract}
We use a ferromagnetic voltage probe model to study the influence
of inelastic scattering on giant magnetoresistance and
current-induced torques in ferromagnetic and antiferromagnetic
metal spin valves. The model is based on the Green's function
formulation of transport theory and represents spin-dependent and
spin-independent inelastic scatterers by interior voltage probes
that are constrained to carry respectively no charge current and
no spin or charge current.  We find that giant magnetoresistance
and spin transfer torques in ferromagnetic metal spin valve
structures survive arbitrarily strong spin-independent inelastic
scattering, while the recently predicted analogous phenomena in
antiferromagnetic metal spin valves are partially suppressed. We
use toy-model numerical calculations to estimate spacer layer
thickness requirements for room temperature operation of
antiferromagnetic metal spin valves.
\end{abstract}
\vskip2pc

\maketitle

\section{Introduction}\label{sec:intro} Electronic phase coherence
often plays an important role in mesoscopic quantum transport. The
simplest example, perhaps, is the Aharonov-Bohm effect
\cite{washburn1986} in a mesoscopic ring which is manifested by
dependence of conductance on enclosed magnetic flux and requires
phase coherence across the ring. In the Landauer-B\"uttiker theory
of quantum transport \cite{landauer1957,buttiker1986} it is
possible to simulate phase breaking scattering processes by
including internal `floating' voltage probes in addition to source
and drain \cite{buttiker1988} electrodes, and requiring that their
chemical potentials adjust so that they do not carry a current.
This simple model is compatible with a Green's function
description of transport and is able to describe those qualitative
consequences of inelastic scattering not associated with the
peculiarities of a specific mechanism.  It has, for example, been
applied to study the influence of phase decoherence on
spin-independent transport in one-dimensional systems
\cite{damato1990, maschke1991}, and the effect of dephasing on
transport through quantum dots \cite{baranger1995, brouwer1995}.

In this paper we report on a comparison of the influence of
inelastic scattering on giant magnetoresistance (GMR) and
spin-transfer (ST) torque effects in traditional spin-valve
structures containing ferromagnetic metals, and in the
antiferromagnetic spin-valve structures that
we\cite{nunez2005,wei2006} have recently proposed. (Spin-valves
are illustrated schematically in Fig.~\ref{fig:fig1}.) These
devices are normally intended for operation at room temperature
and above and electron scattering is therefore normally dominantly
inelastic.  In traditional ferromagnetic spin-valves only
spin-flip scattering in the paramagnetic spacer is expected to
have a large impact on GMR\cite{baibich1988,barnas1990} or
ST\cite{slonczewski1996,berger1996}. Both effects are expected to
be strong when the spin diffusion length is longer than the
paramagnetic spacer thickness.  (Spin-independent inelastic
scattering does however play a role in limiting the amplitude of
oscillations in the dependence of exchange coupling between
ferromagnetic layers on spacer layer thickness \cite{parkin1990}.)
It is commonly believed that GMR and ST in ferromagnetic
spin-valves can survive arbitrarily strong spin-independent
inelastic scattering. For antiferromagnetic spin-valves, however,
the analogous effects depend\cite{nunez2005}, at least in a simple
toy model, on coherent multiple-scattering in the spacer layer and
at its interfaces with the antiferromagnets. We have therefore
argued that both effects will become weak when inelastic
scattering is strong, even if the scattering is spin-independent.
Since phonon scattering is only weakly spin-dependent and strong
at high temperatures, inelastic scattering cannot be ignored in
practical antiferromagnetic spin valves.

To pursue these ideas more quantitatively we use a voltage probe
model to represent the influence of spin-independent and
spin-dependent inelastic scatterers in a Green's function
description of transport through a spin valve. In the linear
response regime the voltage probe model has been shown to be
equivalent to local coupling to a harmonic-oscillator heat bath
\cite{datta1990}. In a magnetic metal circuit, however,
conventional voltage probe models produce both longitudinal and
transverse spin relaxation. Consider for example a largely
collinear magnetic configuration with a natural spin-quantization
axis.  Spin components transverse to this axis are represented in
quantum mechanics by coherence between spin-up and spin-down
projections of the form $|\!\uparrow\rangle +
e^{i\phi}|\!\downarrow\rangle$. A conventional paramagnetic
voltage probe will change the magnitude and randomize the phase of
both $|\!\uparrow\rangle$ and $|\!\downarrow\rangle$ parts of this
spinor independently and therefore alter all spin-density
components.  This property was used very recently by Michaelis and
Beenakker to model spin decay in chaotic quantum dots
\cite{michaelis}.  This feature is however undesirable in
modelling the influence of inelastic scattering on
magnetoresistance properties of spintronic devices, because the
strongest inelastic scatterers are often phonons, and phonons
conserve spin to a good approximation. In our calculations we
therefore use a voltage probe model generalized to the case of
ferromagnetic probes, shown in Fig.~\ref{fig:fig2}.  This
generalization allows us to consider separately spin-dependent and
spin-independent inelastic scatterers.  The spin-independent case
is realized by adjusting the magnetization direction and the
majority and minority spin chemical potentials of the voltage
probes so that the probe carries neither charge or spin currents.
\begin{widetext}
\begin{center}
\begin{figure}
\vspace{-0.5cm} \epsfig{figure=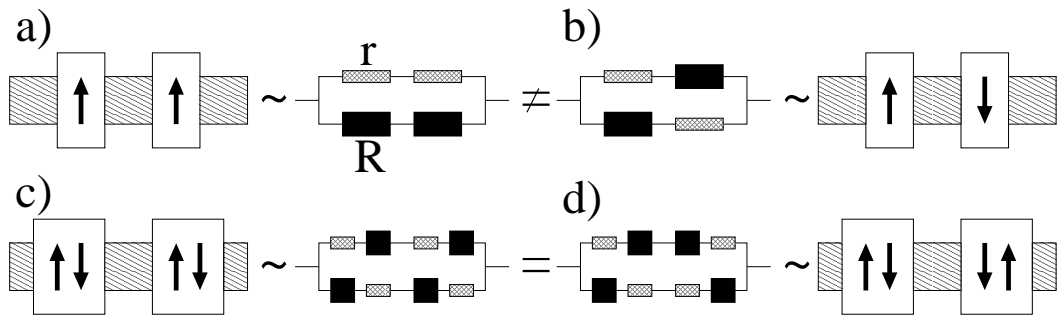, width=15cm}
 \caption{a) Parallel configuration and equivalent circuit consisting of two resistors $R$ and $r<R$, and
 b) antiparallel configuration and equivalent circuit, of a spin valve
 consisting of two single-domain ferromagnets separated by a paramagnetic spacer.
 c) Antiparallel configuration and equivalent circuit of an antiferromagnetic spin valve
 and d) its parallel configuration and equivalent circuit.
 (For convenience, only two ferromagnetic layers within each antiferromagnet are shown.)
 Note that in the ferromagnetic case the resistance of  parallel and antiparallel
 circuits is different, contrary to the equivalent circuits in the antiferromagnetic case. }
 \label{fig:fig1}
\end{figure}
\end{center}
\end{widetext}
There are a number of important distinctions between GMR and ST
physics in ferromagnetic and antiferromagnetic\cite{nunez2005}
structures. In the ferromagnetic case GMR can be understood
qualitatively simply by using a two-channel resistor model
illustrated in Fig.~\ref{fig:fig1} which does not rely on phase
coherence.  When applied to antiferromagnetic spin valves, the
same argument does not predict GMR because the two spin channels
have identical conductances (see Fig.~\ref{fig:fig1}. Clearly the
antiferromagnetic spin-valve GMR effect relies at least in part on
phase-coherence near the spacer layer as. In the simple model we
discuss below antiferromagnetic spin-valve GMR comes about,
roughly speaking, because the reflection amplitude for an electron
scattering off an antiferromagnet is spin dependent, although the
reflection probability and transmission amplitude and probability
are not. Similarly ferromagnetic spin-valve ST torques can be
understood by using a model of a ferromagnetic metal which has
quasiparticle and magnetization orientation degrees of freedom,
and appealing to conservation of total spin angular momentum to
infer an action-reaction relationship between the torque exerted
on the quasiparticles by the exchange field and the torque exerted
on the magnetization by current-carrying quasiparticles.  In the
antiferromagnetic case, it is the staggered moment
antiferromagnetic order which behaves collectively.  Since this
coordinate does not carry total spin, its current-driven dynamics
is not specified by a global conservation law.  The theory of
current-driven order parameter dynamics in this case requires a
more microscopic approach\cite{nunez2004} in which ST is seen as
following from changes in the spin-dependent exchange potential
experienced by all quasiparticles, which follow in turn from
changes in the spin-density in the presence of a transport
current.  Indeed the use of the term {\em spin transfer} torque is
perhaps inappropriate in the antiferromagnetic case since
current-induced order parameter changes {\em are not} related to
transfer of total spin angular momentum between subsystems.
Nevertheless staggered torques {\em do} act on the staggered spins
and {\em do} drive drive the antiferromagnetic order parameter. To
make clear that current-induced order parameter changes in
antiferromagnets are not related to conservation of spin and are
therefore strictly speaking not an example of spin transfer we,
from now on, will call these torques {\em current-induced torques}
(CIT's). Remarkably, current-induced torques in an
antiferromagnetic spin-valve act throughout the volume of the
antiferromagnets in the absence of inelastic scattering and
disorder. Critical currents for order parameter reversal are
therefore independent of antiferromagnetic film thickness in this
limit, rather than being inversely proportional to thickness as in
the ferromagnetic case.

The main aim of this paper is to shed light on the robustness of
these GMR and current-induced torques in realistic
room-temperature thin-film antiferromagnetic spin-valve structures
by examining how properties change upon introduction of
spin-dependent and spin-independent inelastic scatterers.  The
ferromagnetic voltage probe models of elastic and inelastic
scattering are completely satisfactory for this purpose because we
are interested in achieving a qualitative understanding that
transcends the details of specific systems. We solve the model
using the Green's function formalism for electronic transport in
mesoscopic systems \cite{caroli1972,dattabook}. This is convenient
because the spin-torques we want to evaluate can be
expressed\cite{nunez2004} in terms of the transport steady-state
electron spin density and local observables are readily calculated
using the Green's function formalism.

Our paper is organized as follows. In Sec.~\ref{sec:negf} we
describe the Green's function formalism for electronic transport
as applied to systems with ferromagnetic leads. (For related work
see Refs.~[\onlinecite{zhu2003}] and
Ref.~[\onlinecite{yanik2006}].) The main result of this section is
a general expression for the spin current from the ferromagnetic
leads into the system which is necessary to apply our voltage
probe model of a spin-independent inelastic scatterer. In
Sec.~\ref{sec:appls} we apply the formalism to study the effect of
spin-independent and spin-dependent inelastic scattering on
magnetoresistance and current-induced torques in both
ferromagnetic and antiferromagnetic systems.  We conclude that
although GMR and current-induced torques in antiferromagnetic
spin-valves will be weakened by the inelastic scattering always
present at spintronic device operation temperatures, the effects
predicted in Ref.~[\onlinecite{nunez2005}] should still be easily
observable in favorable materials.  We end in Sec.~\ref{sec:concl}
with a discussion of our results and some suggestions for future
work.

\begin{widetext}
\begin{center}
\begin{figure}
\vspace{-0.5cm} \epsfig{figure=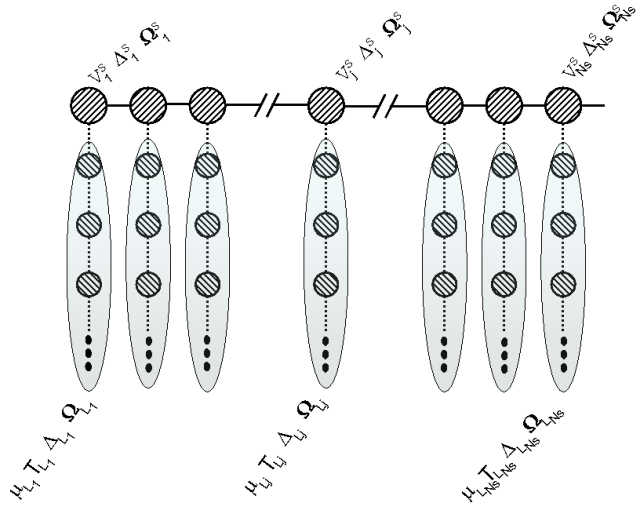, width=10cm}
 \caption{Model system consisting of sites with an arbitrary onsite exchange potential.
 In addition to the coupling to its nearest neighbors, each site is coupled to a ferromagnetic reservoir.}
 \label{fig:fig2}
\end{figure}
\end{center}
\end{widetext}

\section{Green's function formalism with ferromagnetic leads}
\label{sec:negf} In this section we describe our model system and
give an expression for the spin current from the leads into the
system.

\subsection{Model}

We consider first a one-dimensional non-interacting electron
system that is connected on every site $j$ to a ferromagnetic
reservoir at equilibrium with chemical potential $\mu^j$ (see
Fig.~\ref{fig:fig2}). The non-interacting electrons should be
understood as quasiparticles in a mean-field description of a
magnetic metal similar to the Kohn-Sham quasiparticles of
spin-density functional theory. The assumption of
one-dimensionality is not essential. Calculations similar to the
ones we describe which allow for a number of transverse channels
could be carried out to model particular materials systems
realistically.  The generalization of the formalism to two and
three dimensions is straightforward. The total Hamiltonian
$\mathcal{H}=\mathcal{H}_{\rm S}+\mathcal{H}_{\rm
L}+\mathcal{H}_{\rm C} $ is the sum of three parts. The first term
describes the system, which in our case includes the paramagnetic
spacer layer and the magnets. In terms of the second-quantized
operators $\hat c_{j,\sigma}$ that annihilate an electron in spin
state $\sigma \in \{ \uparrow, \downarrow\}$ at site $j$
\begin{eqnarray}
\mathcal{H}_{\rm S} &=& - J_{\rm S} \sum_{\langle
j,j'\rangle;\sigma}
 \hat c_{j,\sigma}^\dagger \hat c_{j',\sigma} \nonumber \\
&+&
 \sum_{j;\sigma,\sigma'} \hat c^\dagger_{j,\sigma}
\left( V^{\rm S}_j \; \delta_{\sigma,\sigma'}- \Delta^{\rm S}_j \; \bm{
\Omega}_j^{\rm S} \cdot  \bm{\tau}_{\sigma,\sigma'} \right)
 \hat c_{j,\sigma'}~.
\end{eqnarray}
In this expression the first term is proportional to the hopping
amplitude $J_{\rm S}$ and its sum is over nearest neighbors only.
In the second and third terms of $\mathcal{H}_{\rm S}$ we allow
for a site-dependent scalar potential $V^{\rm S}_j$ and a
site-dependent exchange potential $-\Delta^{\rm S}_j
\bm{\Omega}^{\rm S}_j$ where $\bm{\Omega}^{\rm S}_j$ is a unit
vector which specifies the (instantaneous) magnetization
orientation on site $j$ and $\bm{\tau}$ is the Pauli spin-matrix
vector.   This single-particle Hamiltonian should be understood as
a time-dependent mean-field Hamiltonian that depends on the
instantaneous $\bm{\Omega}^{\rm S}_j$ values. We assume that the
magnetization dynamics is always slow enough to justify
time-independent quasiparticle transport theory.  The leads are
described by the Hamiltonian $\mathcal{H}_{\rm L} = \sum_j
\mathcal{H}^{{\rm L}_j}$, where $\mathcal{H}^{{\rm L}_j}$ is the
Hamiltonian of the lead that is connected to the system at site
$j$, which is given by
\begin{eqnarray}
\mathcal{H}^{{\rm L}_j} &=& - J_{{\rm L}_j} \sum_{\langle
j',j''\rangle;\sigma}
 \left[ \hat d_{j',\sigma}^{{\rm L}_j} \right]^\dagger \hat d^{{\rm L}_j}_{j'',\sigma} \nonumber \\
&-& \Delta^{{\rm L}_j} \bm{\Omega}^{{\rm L}_j} \cdot
 \sum_{j';\sigma,\sigma'} \left[ \hat d_{j',\sigma}^{{\rm L}_j} \right]^\dagger
\bm{\tau}_{\sigma,\sigma'}
 \hat d^{{\rm L}_j}_{j',\sigma'}~ ,
\end{eqnarray}
where $\hat d^{{\rm L}_j}_{j,\sigma}$ are the fermionic
annihilation operators of the $j$-th lead and $J_{{\rm L}_j}$ is
the hopping amplitude in that lead. Each lead has a chemical potential $\mu_j$
and a uniform exchange potential $-\Delta^{{\rm L}_j} \bm{\Omega}^{{\rm L}_j}$
which we adjust as described earlier to simulate spin-independent and
spin-dependent inelastic scatterers.  Note that $\bm{\Omega}^{{\rm L}_j}$
is in general different for different leads. The Hamiltonian
that couples the system and its leads is $\mathcal{H}_{\rm C} = \sum_j \mathcal{H}^j_{\rm C}$
where
\begin{equation}
 \mathcal{H}^j_{\rm C} = -J^j_{\rm C} \sum_\sigma \left[ \hat c^\dagger_{j,\sigma} \hat d^{{\rm L}_j}_{\partial{\rm L}_j,
 \sigma} + \left[ \hat d^{{\rm L}_j}_{\partial{\rm L}_j,\sigma}\right]^\dagger \hat c_{j,\sigma}
 \right]~.
\end{equation}
Here, $\partial {\rm L}_j$ denotes the last site of the
half-infinite ferromagnetic reservoir connected to site $j$, and
$J_{\rm C}^j$ is the amplitude to hop from that reservoir to the
system.

\subsection{Quantum Transport Green's Function Formalism}

The quantum transport Green's function formalism determines the
equal-time ``lesser" Green's function  \cite{caroli1972,dattabook}
$G^<_{j,\sigma;j',\sigma'} (t,t) \equiv i \langle \hat
c^\dagger_{j',\sigma'} (t) c_{j,\sigma} (t) \rangle$ of the system
in the transport steady state, the quantity in terms of which all
observables are calculated. It is given by
\cite{caroli1972,dattabook}
\begin{equation}
\label{eq:kineq}
  - i G^< (t,t) \equiv  \rho  = \sum_j \int \frac{d\epsilon}{(2\pi)}
 N (\epsilon-\mu^j) A^j (\epsilon)~,
\end{equation}
where $N(x) = (e^{\beta x}+1)^{-1}$ is the Fermi distribution
function, $\beta=1/k_{\rm B}T$ the inverse thermal energy, and
\begin{equation}
 A^j (\epsilon) = G^{(+)} (\epsilon) \hbar \Gamma^j (\epsilon) G^{(-)} (\epsilon)~,
\end{equation}
the spectral weight contribution from lead $j$.  The matrix
elements of the rate $\hbar \Gamma^j (\epsilon)$ are related to
the retarded self energy of lead $j$, denoted by $\hbar
\Sigma^{j,(+)} (\epsilon)$, by
\begin{equation}
 \hbar \Gamma^j (\epsilon) = i\left[ \hbar \Sigma^{j,(+)} (\epsilon)  -  \hbar \Sigma^{j,(-)} (\epsilon)\right]~,
\end{equation}
where $\hbar \Sigma^{j,(-)} (\epsilon)$ is the complex transpose
of the retarded self energy. ($G^{(-)}$ is obtained similarly from
the expression for $G^{(+)}$ below.) The only nonzero elements of
the retarded self-energy are
\begin{eqnarray}
  && \hbar \Sigma^{j, (+)}_{j,\sigma;j,\sigma'}(\epsilon) =
  -\frac{\left(J^j_{\rm C}\right)^2}{2 J_{{\rm L}_j}}\left[ e^{i k^j_\uparrow (\epsilon)a}+ e^{i k^j_\downarrow (\epsilon)
  a}\right]\delta_{\sigma,\sigma'} \nonumber \\
  && \;\;\;\; - \frac{\left(J^j_{\rm C}\right)^2}{2 J_{{\rm L}_j}}\left[ e^{i k^j_\uparrow (\epsilon)a}-e^{i k^j_\downarrow (\epsilon) a}\right]
   \bm{\Omega}^{{\rm L}_j} \cdot \bm{\tau}_{\sigma,\sigma'}~,
\end{eqnarray}
where $k^j_\sigma (\epsilon) =\arccos [-(\epsilon+\sigma
\Delta^{{\rm L}_j})/2 J_{{\rm L }_j}]/a$ is the wave vector in the
leads at energy $\epsilon$, with $a$ the lattice constant.
Finally, the retarded Green's function is specified by
\begin{equation}
\left[  \epsilon^+ - H - \sum_j \hbar \Sigma^{j, (+)} (\epsilon)
\right] G^{(+)} (\epsilon) = 1~,
\end{equation}
with $\epsilon^+ \equiv \epsilon+i0$ where the system Hamiltonian
is
\begin{eqnarray}
  H_{j,\sigma;j',\sigma'} &=& - J_{\rm S} \delta_{\sigma,\sigma'} \left( \delta_{j,j'-1}+ \delta_{j,j'+1}\right) \nonumber \\
  &+& \delta_{j,j'} \left( V^{\rm S}_j \delta_{\sigma,\sigma'}- \Delta_j^{\rm S} \bm{\Omega}^{\rm S}_j \cdot \bm{\tau}_{\sigma,\sigma'} \right)~.
\end{eqnarray}
Note that $H$, $\rho$, $G^<$, $A^j$, $\hbar \Gamma^j$, $\hbar
\Sigma^{j, (\pm)}$ and $G^{(\pm)}$ are matrices in real and spin
space of dimension $(2N_{\rm s}) \times (2N_{\rm s})$, where
$N_{\rm s}$ is the number of sites in the system. From now on we
use the convention that any quantity is a matrix in those indices
that are not explicitly indicated.  For example, the quantity
$G^{(+)}_{i,j} (\epsilon)$ is a $2\times2$ matrix in spin space.

We now proceed to calculate the current and spin currents from the
leads into the system. Since the underlying model Hamiltonian in
these model calculations are spin-rotationally invariant, each
component of spin is conserved and there is no ambiguity in the
spin-current definition.  The expression for the current from lead
$j$ has been derived previously \cite{meir1992} and is given by
\begin{eqnarray}
 - \frac{dN^j}{dt} &\equiv& -\frac{d}{dt} \; \sum_{j',\sigma} \left\langle \left[\hat
 d^{{\rm L}_j}_{j',\sigma} (t) \right]^\dagger \hat d^{{\rm L}_j}_{j',\sigma} (t) \right\rangle \nonumber \\
 &=&\frac{1}{\hbar} \int \frac{d\epsilon}{(2\pi)}
 \sum_k
 \left[ N(\epsilon-\mu^j) -N(\epsilon-\mu^k) \right] \nonumber \\
&&
 {\rm Tr}
 \left[
  \hbar \Gamma^j (\epsilon) G^{(+)} (\epsilon) \hbar \Gamma^k (\epsilon)
  G^{(-)} (\epsilon)
 \right]~.
\end{eqnarray}
This expression relates the Landauer-B\"utikker formalism to the
Green's function formalism \cite{fisher1981}.

The expression for the spin current into lead $j$ proceeds along
similar lines. We first calculate the rate of change of
spin-density in lead $j$:
\begin{eqnarray}
&&  \frac{d {\bf s}^j}{dt}  \equiv  \frac{d }{dt} \;
\sum_{j';\sigma,\sigma'}
 \left\langle \left[\hat
 d^{{\rm L}_j}_{j',\sigma} (t) \right]^\dagger  \frac{{\bm \tau}_{\sigma,\sigma'}}{2} \hat d^{{\rm L}_j}_{j',\sigma'} (t) \right\rangle
\nonumber \\ &&= - \frac{i J^j_{\rm C}}{2 \hbar}
 \sum_{\sigma,\sigma'} \bm{\tau}_{\sigma,\sigma'}
  \left[ \left\langle \rule{0mm}{5mm} \hat c^\dagger_{j,\sigma} (t) \hat d^{{\rm L}_j}_{\partial {\rm L}_j,\sigma'} (t) \right\rangle
 - {\rm c.c.}
 \right].
\end{eqnarray}
In principle, there is an additional term that describes the precession of
electronic spins in the lead around the exchange field in the leads, but
this term vanishes since the leads are assummed to be in equilibrium and
assumed to have a spin that is aligned with the exchange field.

The evaluation of the above expression follows the same line as in
the charge current expression derived by Meir and Wingreen
\cite{meir1992}. We first introduce the Green's function $G^{{\rm
C},<}_{j;\sigma,\sigma'} (t,t') \equiv i \left\langle \left[\hat
d^{{\rm L}_j}_{\partial {\rm L}_j,\sigma'} (t')\right]^\dagger
\hat c_{j,\sigma} (t)\right\rangle$. The corresponding Keldysh
contour ordered Green's function obeys the Dyson equation
\cite{meir1992}
\begin{equation}
  G^{\rm C}_j (t,t') = \frac{J^j_{\rm C}}{\hbar}
  \int_{\mathcal{C}^\infty} dt'' G_{j,j}(t,t'') G^{{\rm L}_j}_{\partial {\rm L}_j, \partial {\rm
  L}_j}(t'',t')~,
\label{eq:dyson}
\end{equation}
where the time integration is over the Keldysh contour
$\mathcal{C}^\infty$.  In Eq.~(\ref{eq:dyson}) $G (t,t')$ is the
contour ordered Green's function of the system coupled to the
leads whereas, in order to avoid double-counting of the effects of
coupling between system and leads, $G^{{\rm L}_j} (t,t')$ is the
contour ordered Green's functions for the $j$-th lead {\em in the
absence} of coupling to the system. Using standard rules for
calculus on the Keldysh contour we find that
\begin{eqnarray}
  G^{{\rm C},<}_j (t,t') &=& \frac{J^j_{\rm C}}{\hbar}
  \int dt'' \Big[ G^{(+)}_{j,j}(t,t'') G^{{\rm L}_j,<}_{\partial {\rm L}_j, \partial {\rm
  L}_j}(t'',t') \nonumber \\
&+& G^{<}_{j,j}(t,t'') G^{{\rm L}_j,(-)}_{\partial {\rm L}_j, \partial {\rm  L}_j}(t'',t') \Big]~,
\end{eqnarray}
which after Fourier transformation results in
\begin{eqnarray}
  G^{{\rm C},<}_j (\epsilon)  &=& J^j_{\rm C}
   \left[ G^{(+)}_{j,j}(\epsilon) G^{{\rm L}_j,<}_{\partial {\rm L}_j, \partial {\rm
  L}_j}(\epsilon) \right. \nonumber \\ && \left.  + G^{<}_{j,j}(\epsilon) G^{{\rm L}_j,(-)}_{\partial {\rm L}_j, \partial {\rm
  L}_j} (\epsilon) \right]~.
\end{eqnarray}
This expression is evaluated using
\begin{equation}
  G^{{\rm L}_j,<}_{\partial {\rm L}_j, \partial {\rm
  L}_j} (\epsilon) = \frac{i}{\left( J^j_{\rm C}\right)^2} N (\epsilon-\mu^j) \hbar
  \Gamma^j_{j,j} (\epsilon)~,
\end{equation}
\begin{equation}
  G^{{\rm L}_j,(-)}_{\partial {\rm L}_j, \partial {\rm
  L}_j} (\epsilon) = \frac{1}{\left(J^j_{\rm C}\right)^2}  \hbar
  \Sigma^{j,(-)}_{j,j} (\epsilon)~,
\end{equation}
and the kinetic equation [Eq.~(\ref{eq:kineq})] to obtain
\begin{eqnarray}
  G^{{\rm C},<}_j (\epsilon) &=& \frac{i}{J^j_{\rm C}}
  \Big[ N (\epsilon-\mu^j) G^{(+)}_{j,j} (\epsilon) \hbar \Gamma^j_{j,j} (\epsilon) \nonumber \\
&+&   \sum_k N (\epsilon-\mu^k) A^k_{j,j} (\epsilon) \hbar
  \Sigma^{j,(-)}_{j,j}\Big].
\end{eqnarray}
It follows that
\begin{widetext}
\begin{equation}
\label{eq:spincurrentresult}
 \frac{d {\bf s}^j}{dt}
 = \frac{1}{2 \hbar} \int \frac{d\epsilon}{(2\pi)}
  {\rm Tr} \left\{ \rule{0mm}{6mm}
    N (\epsilon-\mu^j) \hbar \Gamma^j_{j,j} (\epsilon) i \left[ \bm{\tau} G^{(+)}_{j,j} (\epsilon) - G^{(-)}_{j,j} (\epsilon) \bm{\tau} \right]
   \right.
   \left. - \sum_k N (\epsilon-\mu^k) i \left[\bm{\tau} \hbar \Sigma^{j,(+)}_{j,j} (\epsilon) - \hbar \Sigma^{j,(-)}_{j,j} (\epsilon) \bm{\tau} \right]
  A^k_{j,j} (\epsilon)
  \right\}~.
\end{equation}
\end{widetext}
In the next subsection we turn to a discussion of the physical
content of this equation.

\subsection{Spin-Currents, Exchange Interactions, and Current-Induced Torques}

\label{subsec:stt} In order to provide some insight into the
general expression we have derived for the spin-current, we examine
first the relatively simple situation in which the system is connected to leads
only on the most left and most right sites, denoted by site $1$ and $N_{\rm s}$
respectively. Their respective chemical potentials are $\mu^1 =
\epsilon_{\rm F} + |e|V$, and  $\mu^{N_{\rm s}} =\epsilon_{\rm F}
$ with $|e|V>0$. Moreover, we assume that there is no exchange
potential in the system, i.e., $\Delta^{\rm S}_j=0$, so that spin
currents are conserved in the system.  This is the circumstance then of
transport between ferromagnetic leads through a paramagnetic system.
In this case $d {\bf s}^1/dt =- d {\bf s}^{N_{\rm s}}/dt$.  It is
informative to separate the spin-current flowing between leads into
equilibrium and non-equilibrium contributions:
\begin{equation}
  \frac{d {\bf s}^1}{dt} =  \left. \frac{d {\bf s}^1}{dt} \right|_{{\rm eq}} +
  \left. \frac{d {\bf s}^1}{dt} \right|_{\rm neq}~.
\end{equation}
It follows quite generally from the time-dependent Schroedinger
equation satisified by quasiparticles that $d {\bf s}/dt$ has contributions
from the spin-current divergence and from precession around an effective magnetic
field.  Since the leads are by definition spatially homogeneous, it follows that
spin-currents in the leads may be equally well thought of as a torque acting on the leads.
The equilibrium spin-torque is given (at temperatures low compared to Fermi energies) by
\begin{widetext}
\begin{equation}
\left. \frac{d {\bf s}^1} {dt} \right|_{{\rm eq}}
 = \frac{1}{2 \hbar} \int^{\epsilon_{\rm F}}_{-\infty} \frac{d\epsilon}{(2 \pi)}
 {\rm Tr} \left\{ \rule{0mm}{6mm}
  \hbar \Gamma^1_{1,1} (\epsilon) i \left[ \bm{\tau} G^{(+)}_{1,1} (\epsilon) -
   G^{(-)}_{1,1} (\epsilon) \bm{\tau} \right] \right. \nonumber \\
    \left. -
   i \left[ \bm{\tau} \hbar \Sigma^{1,(+)}_{1,1} (\epsilon) - \hbar \Sigma^{1,(-)}_{1,1} (\epsilon)
   \bm{\tau}\right] A_{1,1}(\epsilon) \rule{0mm}{6mm}
 \right\}~,
\end{equation}
where $A(\epsilon) \equiv i \left[ G^{(+)} (\epsilon) - G^{(-)}
(\epsilon) \right]$ is the total spectral function. It can be
shown that the equilibrium torque always points out of the plane
spanned by $\bm{\Omega}^{{\rm L}_1}$ and $\bm{\Omega}^{{\rm
L}_{N_{\rm s}}}$, the magnetization directions of the two leads.
This is expected, since the equilibrium contribution to the
torques between the leads is a simply due to exchange coupling
mediated by the paramagnetic system.  The dynamics induced by this
coupling conserves total energy and hence leaves
$\bm{\Omega}^{{\rm L}_1} \cdot  \bm{\Omega}^{{\rm L}_{N_{\rm s}}}$
invariant. Note that spin currents are even under time reserval,
unlike charge currents, and can be present in equilibrium.

The nonequilibrium (transport) contribution to the torques between
the two leads, is given by
\begin{equation}
\left. \frac{d {\bf s}^1}{dt} \right|_{\rm neq}
 = \frac{1}{2\hbar} \int_{\epsilon_{\rm F}}^{\epsilon_{\rm F}+ |e|V}
 \frac{d\epsilon}{(2\pi)}
  {\rm Tr} \left\{ \rule{0mm}{6mm}
    \hbar \Gamma^1_{1,1} (\epsilon) i \left[\bm{\tau} G^{(+)}_{1,1} (\epsilon) -
     G^{(-)}_{1,1} (\epsilon) \bm{\tau} \right]
   \right.
   \left. - i \left[ \bm{\tau} \hbar \Sigma^{1,(+)}_{1,1} (\epsilon) - \hbar \Sigma^{1,(-)}_{1,1} (\epsilon) \bm{\tau} \right]
  A^{1}_{1,1} (\epsilon)
  \rule{0mm}{6mm} \right\}~,
\end{equation}
\end{widetext}
and has both an out-of-plane and in-plane contribution. The former
corresponds to a transport-modified electron-mediated exchange
torque between the leads. The latter corresponds to the spin
transfer torque between the ferromagnetic leads.  (This
contribution was also calculated within the context of the Green's
function formalism in Ref.~[\onlinecite{zhu2003}] for the case of
two ferromagnetic leads.) Note that the in-plane spin transfer
torque is present only at finite bias $|e|V$, which reflects the
fact that energy-conservation violating torques can only be
present in a nonequilibrium situation. Finally, note that this
spin transfer torque acts on the leads and should not be confused
with the torques that act on the local magnetization in the system
that we calculate in the next section.

\section{Applications}
\label{sec:appls}

The ferromagnetic voltage probe model discussed in the previous
section is a flexible tool which can be used to model a wide variety of
potentially interesting spintronic device geometries.  We use
it in this section to model both spin-independent and
spin-dependent inelastic scattering.

In the presence of an exchange field which defines a preferred
direction in spin space, we can recognize (at least) three
different length scales that are relevant to spintronic device
functionality.  The phase-coherence length $L_\phi$ is the length
over which electronic quasiparticles maintain phase coherence. The
spin-flip scattering length or spin diffusion length $L_{\rm sf}$
is the average distance travelled along the channel between
spin-flip scattering events.  This quantity usually controls the
length scale on which the magnititude of the magnetization along
the exchange field direction recovers local equilibrium, and
therefore also the paramagnetic spacer layer thickness scale at
which giant magnetoresistance effects are strongly attenuated. The
spin-coherence length $L_{\rm sc}$ is the length scale over which
components of the spin-density transverse to the exchange field
can be maintained.  The phase-coherence length is limited by
spin-dependent inelastic scattering processes. In addition to this
inelastic contribution, the spin-coherence length in two and three
dimensional ferromagnetic metals tends to be short because of
destructive interference due to the phase difference of the spinor
components in transverse conduction channels\cite{stiles2002} with
different Fermi velocity components in the current flow direction.
As we discuss below, phase coherence can have a strong influence
on the properties of a spintronic circuit, especially in circuits
containing antiferromagnetic elements.

The solution of transport Green's function models with
ferromagnetic leads and non-collinear magnetization presents a
number of numerical challenges, even for the one-dimensional case
we consider here, because the spin and charge currents in the
leads depend nonlinearly on both the magnetization direction and
the magnitude of the exchange spin-splitting in each lead.  When
we wish to model a spin-independent inelastic scatterer, both the
magnitude and the direction in each lead have to be carefully
adjusted to achieve the zero-spin-current condition.  Indeed the
case of spin-independent inelastic scatterers can be very relevant
to experiment since phonon-scattering can be dominant in
experimental systems. To model circumstances in which phonon
scattering is dominant ($L_{\rm sf} \gg L_\phi$), we must deal
with the complications associated with preventing spin-relaxation
in the voltage probes, a requirement that is especially
troublesome for non-collinear magnetization configurations.
Inelastic scattering off magnons, which does flip spins, can also
be important however. The simplest case to examine numerically is
one with paramagnetic, {\em i.e.} spin-isotropic, voltage probes.
Calculations with this voltage probe model are relatively simple
even for non-collinear magnetizations, since in this case only the
charge current into the probes is set to zero. This limit
corresponds to $L_{\rm sf} = L_\phi$ and can be interpreted as
representing the case in which quasiparticle scattering from
magnons is dominant.

We start by considering the
giant magnetoresistance (GMR) ratio which is easier to evaluate because we
need to consider only parallel and antiparallel configurations
which are both collinear. We define the GMR ratio as
\begin{equation}
  \eta = \frac{G_{\rm P}-G_{\rm AP}}{G_{\rm P}}~,
\end{equation}
where $G_{\rm P}$ is the conductance for the parallel
configuration and $G_{\rm AP}$ for the antiparallel one. For
antiferromagnets, the designations {\em parallel} and {\em antiparallel} refer to the
moment directions on the two sites adjacent to the paramagnetic spacer.

In our model we define the phase coherence length as the
product of the Fermi velocity and the inelastic scattering rate:
\begin{equation}
  L_\phi = \frac{4 J_{\rm S} a \sin \left( k_{\rm F} a \right)}
  {\hbar {\rm Tr} \left[ \sum_j\Gamma^j (\epsilon_{\rm F})/N_{\rm s} \right]}~,
\end{equation}
where $k_{\rm F}$ is the Fermi wave length of the system. This
definition is motivated the fact that the inelastic scattering
time is given roughly by $2/{\rm Tr} \left[ \sum_j \Gamma^j
(\epsilon_{\rm F})/N_{\rm s} \right]$.

\begin{figure}
\vspace{-0.5cm}
\begin{center}\epsfig{figure=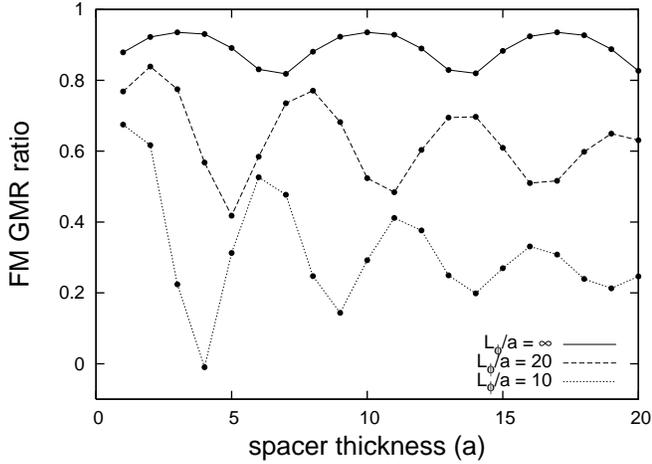}
 \caption{Ferromagnetic GMR ratio as a function of the spacer thickness
 for various inelastic scattering lengths.}
 \label{fig:fm_gmr}\end{center}
\end{figure}

\begin{figure}
\vspace{-0.5cm}
\begin{center}\epsfig{figure=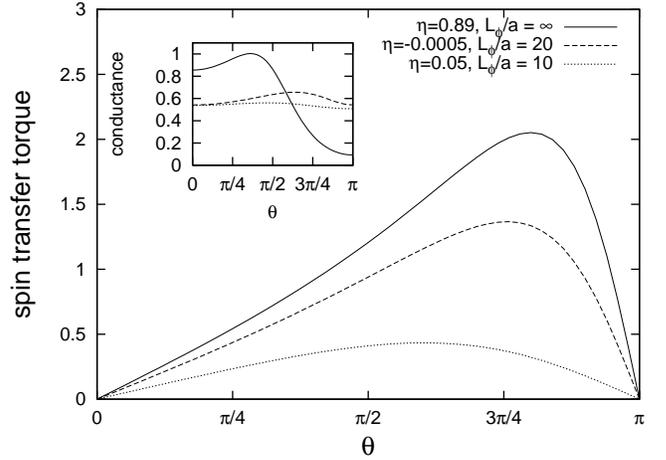}
 \caption{Spin transfer torque for the case $L_{\rm sf} = L_\phi$
 normalized to $I/|e|$, where $I$ is the charge current, as a function of $\theta$.
 The inset shows the conductance in units of $|e|/(2\pi\hbar)$.
 For this calculation the spacer thickness is  taken equal to $5$ sites.}
 \label{fig:sttfm} \end{center}
\end{figure}

\subsection{Ferromagnetic Metal Nanostructures}
\label{subsec:fms} We report results for a spin-valve
ferromagnetic nanostructure, which consists of two ferromagnetic
elements separated by a paramagnetic spacer (See
Fig.~\ref{fig:fig1}).  This structure is known as a spin-valve
because the current can be sharply reduced under favorable
circumstances when the two ferromagnets have opposite
orientations. In our toy model we choose $\Delta^{\rm S}_j/J_{\rm
S}=0.5$ and independent of position, {\em i.e.} the site index
$j$, in the ferromagnetic parts of the system. In the paramagnetic
parts the exchange potential is set to zero. Both ferromagnets are
$4$ sites long in our tight-binding model. The other parameters
used for the calculations reported on here were $\epsilon_{\rm
F}/J_{\rm S }=1.8$, $J_{\rm S}=J_{{\rm L}_j}$ and $V^{\rm S}_j$.

\subsubsection{GMR ratio for $L_{\rm sf} \gg L_{\rm \phi}$}
To model a finite phase coherence length due to spin independent
inelastic scattering, the phonon scattering case, we choose
$L_{\rm sf}=\infty$. We then have to require that both spin and
charge currents into the voltage probes vanish. As explained
before, considering the non-collinear case for this situation is
very difficult numerically. However, calculation of the GMR ratio
requires comparing parallel and antiparallel configuration in
which the spin currents with polarization out of the plane are
zero. This observation makes calculating the GMR ratio numerically
tractable.

In Fig.~\ref{fig:fm_gmr} we present results for the GMR ratio as a
function of spacer thickness for various inelastic scattering
lengths. The GMR ratio exhibits oscillations as a function of
spacer thickness which have been observed experimentally
\cite{parkin1990}. For the phase coherent case these oscillations
persist up to arbitrary large spacer thickness in our
one-dimensional model. For finite $L_\phi$ the oscillations are
damped and the GMR ratio saturates to a nonzero value for a spacer
of thickness much larger than $L_\phi$. This is expected because
for a ferromagnet the transmission and reflection probabilities
are spin dependent, and not only the amplitudes.

\subsubsection{GMR ratio and spin transfer torques for $L_{\phi}=L_{\rm sf}$}
For the magnon scattering case, $L_\phi = L_{\rm sf}$, implemented by
making the floating voltage probes paramagnetic and requiring the
charge current to be zero, we are also able to consider the
non-collinear case. The spin transfer torques are due to the net
misalignment of electron spins with the local magnetization in
the transport steady state \cite{nunez2004}, and calculated from
\begin{equation}
  \Gamma^{\rm st} \equiv \sum_j \frac{\Delta^{\rm S}_j}{\hbar} \bm{\Omega}^{\rm S}_j \times \langle {\bf s}
  \rangle_{\perp}~.
\end{equation}
Here, $\langle {\bf s} \rangle_{\perp}$ denotes the electron spin
component that points out of the plane spanned by the two
magnetizations. The sum over sites in the above expression is
restricted to the ferromagnet for which we want to calculate the
spin transfer torques, which in our case is by definition the
right ``downstream"  ferromagnet. In Fig.~\ref{fig:sttfm} the spin
transfer torque per current is shown for various $L_\phi$ as a
function of the angle $\theta$ between the two ferromagnets. The
inset of this figure shows the conductance as a function of this
angle. Clearly, both GMR and spin transfer torques are suppressed
for decreasing $L_\phi$, as expected in the $L_{\rm sf} = L_\phi$
case.

\subsection{Antiferromagnetic Metal Nanostructures}
\label{subsec:afms} In this subsection, we consider a system which
consists of two antiferromagnets separated by a paramagnetic
spacer. Hence, we take $\Delta^{\rm S} _j =\Delta (-1)^{j}$ in the
magnetic parts of the system, and zero in the paramagnetic parts.
We take the parameters $\Delta/J_{\rm S}=1$. The other parameters
used to obtain the results reported on here were $\epsilon_{\rm
F}/J_{\rm S }=1.8$, $J_{\rm S}=J_{{\rm L}_j}$, and $V^{\rm S}_j$.
The calculations presented in this subsection were performed on
nanostructures in which each antiferromagnets has $30$ sites.

\begin{figure}
\vspace{-0.5cm} \centerline{\epsfig{figure=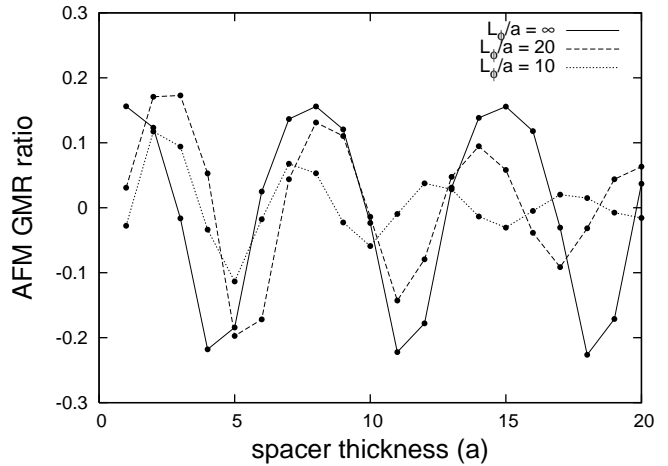}}
 \caption{Antiferromagnetic GMR ratio as a function of the spacer thickness
 for various inelastic scattering lengths.}
 \label{fig:afm_gmr}
\end{figure}

\subsubsection{GMR ratio for $L_{\rm sf} \gg L_{\phi}$}
In Fig.~\ref{fig:afm_gmr} the antiferromagnetic GMR ratio is shown
as a function of the thickness of the spacer, for various
inelastic scattering length $L_\phi$. Note that as the phase
coherence length decreases, the oscillations in the GMR ratio with
increasing thickness are suppressed. Unlike the case of a
ferromagnetic spin valve, the GMR ratio goes to zero when the
thickness of the spacer is much larger than the phase coherence
length.  This property demonstrates that phase coherence is
essential for the antiferromagnetic GMR effect.  This observation
is in agreement with qualitative arguments presented in previous
work \cite{nunez2005}. Note however, that the GMR decreases rather
gradually with decreasing phase coherence length.

\subsubsection{GMR and current-induced torques for $L_{\phi}=L_{\rm sf}$}
One of the most important results in our previous work on
antiferromagnets \cite{nunez2005} is that in the ballistic limit
the out of plane spin density which is responsible for the
current-induced torque is periodic with the period of the
antiferromagnet. For the tight-binding model used in this paper
this property implies that this spin density component, present
only in the transport steady state, is constant throughout the
antiferromagnet. It is this property which makes current-induced
torques very efficient in driving collective order parameter
dynamics in uncompensated antiferromagnets in the absence of
inelastic scattering.

\begin{figure}
\vspace{-0.5cm} \centerline{\epsfig{figure=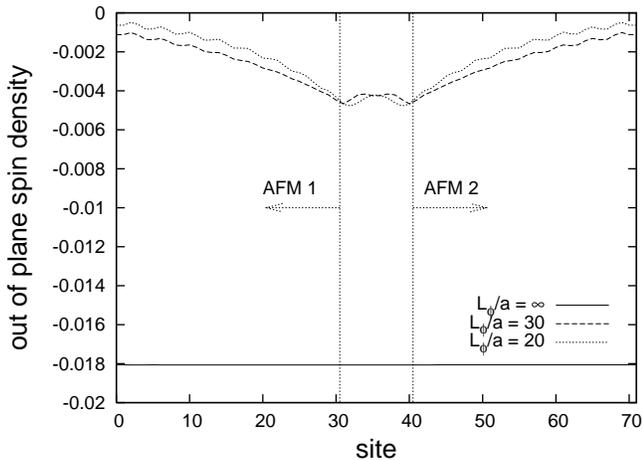}}
 \caption{Spin density component out of the plane spanned by the moments
 of the antiferromagnets opposite the paramagnetic spacer layer for various
 inelastic scattering lengths. In this calculation $L_\phi=L_{\rm sf}$.}
 \label{fig:soutofplane}
\end{figure}

In Fig.~\ref{fig:soutofplane} we show the out of plane spin
density for the angle $\theta=\pi/2$ between two moments in the
antiferromagnets opposite the spacer. This calculation was
performed for a spacer thickness of $10$ sites. In agreement with
our previous results we find that for $L_\phi=\infty$ the out of
plane spin density is constant. For nonzero $L_\phi$ the out of
plane spin density decays away from the spacer into the
antiferromagnets.

\begin{figure}
\vspace{-0.5cm} \centerline{\epsfig{figure=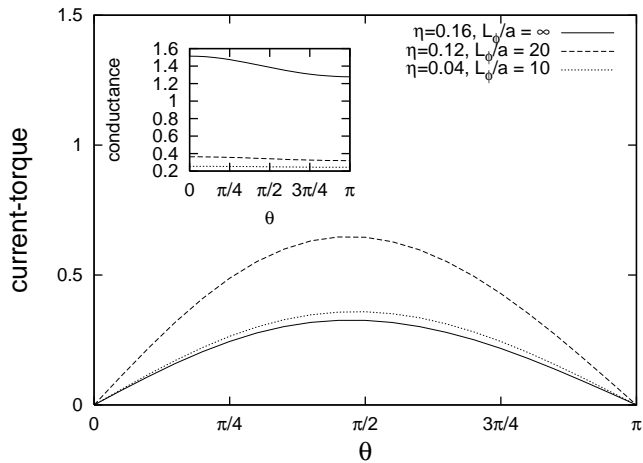}}
 \caption{Current-induced torques normalized to $I/|e|$, where $I$ is the charge current, as a function of $\theta$.
 The inset shows the conductance in units of $|e|/(2\pi\hbar)$.
 In this calculation $L_\phi=L_{\rm sf}$ and the number of spacer sites is $4$.}
 \label{fig:sttafm}
\end{figure}

In Fig.~\ref{fig:sttafm} we show the current-induced torque per
current acting on the right antiferromagnet as a function of the
angle $\theta$ between the two moments facing the paramagnetic
spacer. It is important to realize that a single domain
antiferromagnet is driven by a staggered torque. Hence, in this
case we have to calculate the staggered torque given by
\begin{equation}
  \Gamma^{\rm st} \equiv \frac{1}{\hbar} \sum_j (-1)^j \Delta \bm{\Omega}^{\rm S}_j \times \langle {\bf s}
  \rangle_{\perp}~.
\end{equation}
As expected, both GMR and current-induced torques are reduced by
inelastic scattering. It is somewhat surprising that the
current-induced torques at $L_\phi=20$ are larger than at
$L_\phi=\infty$. However, this is most likely because the
current-induced torque is normalized to the current and because of
the one-dimensional model under consideration. We checked that in
the limit $L_{\phi} \to 0$ the current-induced torques always
vanish.

\section{Discussion and conclusions} \label{sec:concl} The most
important conclusion of this paper is that the introduction of
inelastic scattering does not immediately destroy magnetoresistive
and current-induced torques in nanostructures with two
antiferromagnetic elements separated by a paramagnetic spacer. On
the contrary, we find that the GMR ratio goes to zero smoothly for
spacer thicknesses larger than the phase coherence or inelastic
scattering length. (Ab-initio calculations for Cr/Au/Cr
antiferromagnetic spin-valves\cite{haney2006} demonstrate that
there is an additional interface related contribution to
antiferromagnetic GMR which is not captured by our toy model and
is not limited by inelastic scattering in the spacer layer.) Since
typical inelastic scattering lengths at room temperature ($\sim
10{\rm nm}$) are much larger than the minimum paramagnetic spacer
layer thicknesses required to make coupling between magnetic
layers insignificant, inelastic scattering does not impose any
practical limitation on GMR and current-induced torques in thin
film based antiferromagnetic metal nanostructures.

Unlike the ferromagnetic case, current-induced torques in
antiferromagnets do not follow simply from the approximate
conservation of total spin angular momentum.  Toy model
calculations, as well as {\em ab initio} calculations for
realistic systems\cite{haney2006}, suggest that both GMR and
current-induced torques tend to be somewhat weaker for very thin
films in the antiferromagnetic case compared to GMR and spin
transfer torques in ferromagnets. It seems natural to associate
this property with the absence of grounding in a simple robust
conservation law. A surprising and interesting property of ST
physics in antiferromagnets is the property that torques act
throughout the entire volume of the antiferromagnet in the phase
coherent case.  This property helps to compensate for the tendency
toward somewhat weaker effects in very thin films. When inelastic
scattering is introduced, spin-torques again act only over a
finite thickness $\sim L_{\rm \phi}$ within the antiferromagnets.
Still the length scale over which the torque acts should be $\sim
10{\rm nm}$, much longer than the atomic length scale over which
spin torques act in ferromagnetic spin-torque structures.

As explained in earlier work\cite{nunez2005}, ST effects in
antiferromagnets are expected to be strongest when moment
directions alternate in the direction of current flow.  In the
one-dimensional single channel model studied here, this is the
only possibility.  For real three-dimensional antiferromagnets
this property requires particular orientations of current with
respect to the crystal axes, or in thin film structures particular
growth directions for the antiferromagnetic films which yield
alternating ferromagnetic layers.  It seems likely that the
greatest obstacle to achieving interesting GMR and current-induced
torques in purely antiferromagnetic circuits may be the difficulty
in growing antiferromagnetic thin films with individual
ferromagnetic layers that are not strongly compensated.  Studies
of the influence of magnetic disorder within ferromagnetic layers
on GMR and ST physics require multi-channel Green's function
calculations and will be the subject of future work.

This work was supported by the National Science Foundation under
grants DMR-0606489, DMR-0210383, and PHY99-07949, by a grant from
Seagate Corporation, and by the Welch Foundation. ASN was
supported in part by Proyecto Mecesup FSM0204.

\end{document}